\documentclass[aps,prb,twocolumn,groupedaddress]{revtex4-1}
\usepackage{graphicx}

\begin{document}


\title{Detection of Orbital Fluctuations Above the Structural Transition Temperature in the Iron Pnictides and Chalcogenides}


\author{H. Z. Arham,$^1$ C. R. Hunt,$^1$ W. K. Park,$^1$ J. Gillett,$^2$ S. D. Das,$^2$ S. E. Sebastian,$^2$ Z. J. Xu,$^3$ J. S. Wen,$^3$ Z. W. Lin,$^3$ Q. Li,$^3$ G. Gu,$^3$ A. Thaler,$^4$ S. Ran,$^4$ S. L. Bud'ko,$^4$ P. C. Canfield,$^4$ D. Y. Chung,$^5$ M. G. Kanatzidis,$^5$ L. H. Greene$^1$
}
\affiliation{\\\textsuperscript{$1$}Department of Physics and the Frederick Seitz Material Research Laboratory, University of Illinois at Urbana-Champaign, Urbana, Illinois 61801, USA
\\\textsuperscript{$2$}Cavendish Laboratory, J. J. Thomson Ave, University of Cambridge, UK
\\\textsuperscript{$3$}Condensed Matter Physics and Materials Science Department, Brookhaven National Laboratory, Upton, New York, 11973, USA
\\\textsuperscript{$4$}Ames Laboratory $\&$ Department of Physics and Astronomy, Iowa State University, Ames, IA, 50011, USA
\\\textsuperscript{$5$}Materials Science Division, Argonne National Laboratory, Argonne, IL 60439, USA}


\date{\today}

\begin{abstract}
We use point contact spectroscopy (PCS) to probe $\rm{AEFe_2As_2}$ ($\rm{AE = Ca, Sr, Ba}$) and $\rm{Fe_{1+y}Te}$. For $\rm{AE = Sr, Ba}$ we detect orbital fluctuations above $T_S$ while for AE = Ca these fluctuations start below $T_S$. Co doping preserves the orbital fluctuations while K doping suppresses it. The fluctuations are only seen at those dopings and temperatures where an in-plane resistive anisotropy is known to exist. We predict an in-plane resistive anisotropy of $\rm{Fe_{1+y}Te}$ above $T_S$. Our data are examined in light of the recent work by W.-C. Lee and P. Phillips (arXiv:1110.5917v2). We also study how joule heating in the PCS junctions impacts the spectra. Spectroscopic information is only obtained from those PCS junctions that are free of heating effects while those PCS junctions that are in the thermal regime display bulk resistivity phenomenon. 
\end{abstract}


\maketitle

\section{Introduction}
The parent compounds of the iron based superconductors undergo a $C_4$ symmetry breaking transition when entering an antiferromagnetic orthorhombic ground state.\cite{Cruz, Rotter, Lester, Pratt, Bao, NiNi2} It is not clear if this transition is driven by magnetic fluctuations \cite{Fang, Yildirim} or orbital ordering. \cite{Kruger, Chen1,*Chen2, Lv1,*Lv2, Lee} There is evidence that the quantum critical fluctuations associated with this transition are nematic in character and extend into the normal state of these compounds. \cite{Chu, Harriger}

Point contact spectroscopy (PCS) was discovered in 1974 by I. K. Yanson, \cite{Yanson1} who found nonlinearities in the conductance across a Pb planar tunnel junction where the Pb was driven normal by applied magnetic field. These nonlinearities strikingly mimicked the Eliashberg function, $\alpha^2 F(\omega)$, observed in planar tunnel junctions on Pb by McMillan and Rowell in 1965. \cite{McMillan, Rowell} This observation was surprising because tunneling in the normal state, according to Harrison's theorem, \cite{Harrison} always yields a constant conductance. (Harrison's theorem shows that when tunneling into a Fermi liquid at low bias, the Fermi velocity divides out the density of states). Yanson and co-workers showed that their observed nonlinearities arose from nano-shorts, or ``point contacts'' through the tunnel barrier.  The measurement, therefore, was not tunneling, but quasiparticle scattering; hence PCS is also called quasiparticle scattering spectroscopy (QPS).   

It is easy to understand what is detected in PCS or QPS through a simple, non-quantum mechanical picture. There are three size regimes of a metallic junction: ballistic, also called the Sharvin limit, where the junction is smaller than the electron scattering length; diffusive where the junction size is between the elastic and inelastic scattering lengths; and thermal, where the junction is larger than the electron mean free path. There are many good reviews describing these regimes, \cite{WKPreview, ROPP, Daghero, Duif} so we simply point out the basics here. In the thermal regime, the junction acts like a simple resistor and spectroscopic information cannot be derived. In the ballistic regime, electrons are injected a scattering length into the bulk of the sample, and the Eliashberg function is detected when the electron is inelastically backscattered into the orifice: scattering is detected as a slight decrease in the conductance due to the backflow of electrons. This is a small effect and is detected as peaks in the second harmonic. In the diffusive regime, there is some spectroscopic information, but depending on how close or far the junction is from the ballistic/thermal limit, the spectra can exhibit a range of smearing, which we quantify later in this paper. We also point out an important diagnostic: observing signatures apparent in the resistivity (such as phase transitions) can indicate that the junction is in the thermal limit.

The theory for PCS as a spectroscopic technique for quasiparticle scattering off excitation modes (e.g. phonons, magnons) has been well developed. \cite{Naidyuk, Duif} For single-band s-wave superconductors, the theory is also well established. The seminal work of Blonder, Tinkham, and Klapwijk (BTK) \cite{Blonder} shows how to map out the superconducting density of states from the tunneling to the metallic (Andreev reflection) \cite{Andreev} limits. We stress that the data obtained in the tunneling and Andreev limits look completely different, but with the correct BTK analysis, one can obtain the same spectroscopic results: the gap, phonons, and in the case of extending the BTK theory, the order parameter symmetry. \cite{Tanaka, WKPark}  

However, there is a lack of theoretical work on explaining PCS spectra on novel, strongly correlated materials with non-trivial electron matter. There is no Fermi golden rule type of theory indicating that an increased density of states would yield a larger junction conductance. Nowack and Klug \cite{Nowack} did show how it is possible to explain energy-dependent density of states (DOS) with standard first-order ballistic point contact theory, but as they state, ``a theory treating electronic DOS and scattering as interconnected would be preferable.'' 

Recent work has shown that the PCS technique detects strong electron correlations. For heavy fermion compounds, the onset of the Kondo lattice appears as a Fano lineshape in the PCS spectra. \cite{WKPark} PCS is also sensitive to the hybridization gap in the heavy fermion $\rm{URu_2Si_2}$. \cite{WKPark2} In this work, we show that an increase in the zero bias conductance in the Fe chalcogenides and pnictides can be associated with an increase in the single-particle density of states arising from the onset of orbital ordering fluctuations. \cite{Weicheng} This not only shows that these materials do indeed exhibit such ``electron matter'' but also that PCS is a powerful bulk probe of such states. 

PCS results reported on the Fe-based compounds thus far focus on their superconducting phase. \cite{Daghero, ROPP} We use PCS to measures the differential conductance $dI/dV$ = $G(V)$ of the following compounds at both superconducting and non-superconducting dopings and temperatures: $\rm{AEFe_2As_2}$ (AE = Ca, Sr, Ba), $\rm{Ba(Fe_{1-x}Co_x)_2As_2}$, $\rm{Ba_{0.8}K_{0.2}Fe_2As_2}$ and $\rm{Fe_{1+y}Te}$. For $\rm{Fe_{1+y}Te}$, $\rm{SrFe_2As_2}$ and underdoped $\rm{Ba(Fe_{1-x}Co_x)_2As_2}$ we detect a conductance enhancement well above the magnetic ($T_N$) and structural ($T_S$) transition temperatures; for $\rm{CaFe_2As_2}$ this enhancement is only observed below $T_N$ and $T_S$ while it is not observed at all for $\rm{Ba_{0.8}K_{0.2}Fe_2As_2}$ and overdoped $\rm{Ba(Fe_{1-x}Co_x)_2As_2}$. We relate the $dI/dV$ spectrum to the non-Fermi liquid behavior in these compounds that is manifested as an in-plane resistivity anisotropy, which has been attributed to orbital fluctuations. \cite{Weicheng} 

\section{Experimental Results}
Single crystals of $\rm{SrFe_2As_2}$ and $\rm{Ba(Fe_{1-x}Co_x)_2As_2}$ are grown out of FeAs flux as described in \cite{Sebastian, Gillett} (for x = 0 to 0.08) and \cite{Ni} (for x = 0.085 to 0.125). $\rm{CaFe_2As_2}$ crystals are grown from both Sn and FeAs flux,\cite{Ran} while $\rm{Ba_{0.8}K_{0.2}Fe_2As_2}$ crystals are grown in Sn flux. \cite{Duck} $\rm{Fe_{1+y}Te}$ single crystals are grown by a horizontal unidirectional solidification method. \cite{GGu} Metallic junctions are formed on freshly cleaved c-axis crystal surfaces and $G(V)$ across each junction is measured using a standard four-probe lock-in technique. PCS is carried out in two different configurations: the needle-anvil method and the soft PCS method. \cite{Naidyuk, Daghero} In the needle-anvil setup, an electrochemically polished Au tip is brought into gentle contact with the sample. For soft PCS, we sputter 50{\AA} $\rm{AlO_x}$ on our crystals to act as an insulating barrier. Using Ag paint as a counter electrode, parallel, nanoscale channels are introduced for ballistic current transport by fritting \cite{Holm} across the oxide layer. Similar spectra are obtained from both methods, as is shown in Fig. 4d. In all other figures, raw $G(V)$ data obtained via soft PCS is presented. Soft PCS has the advantage of being more stable with temperature change while needle-anvil PCS gives more control over the junction resistance.  

\begin{figure*}[thbp]
		\includegraphics[scale=0.7]{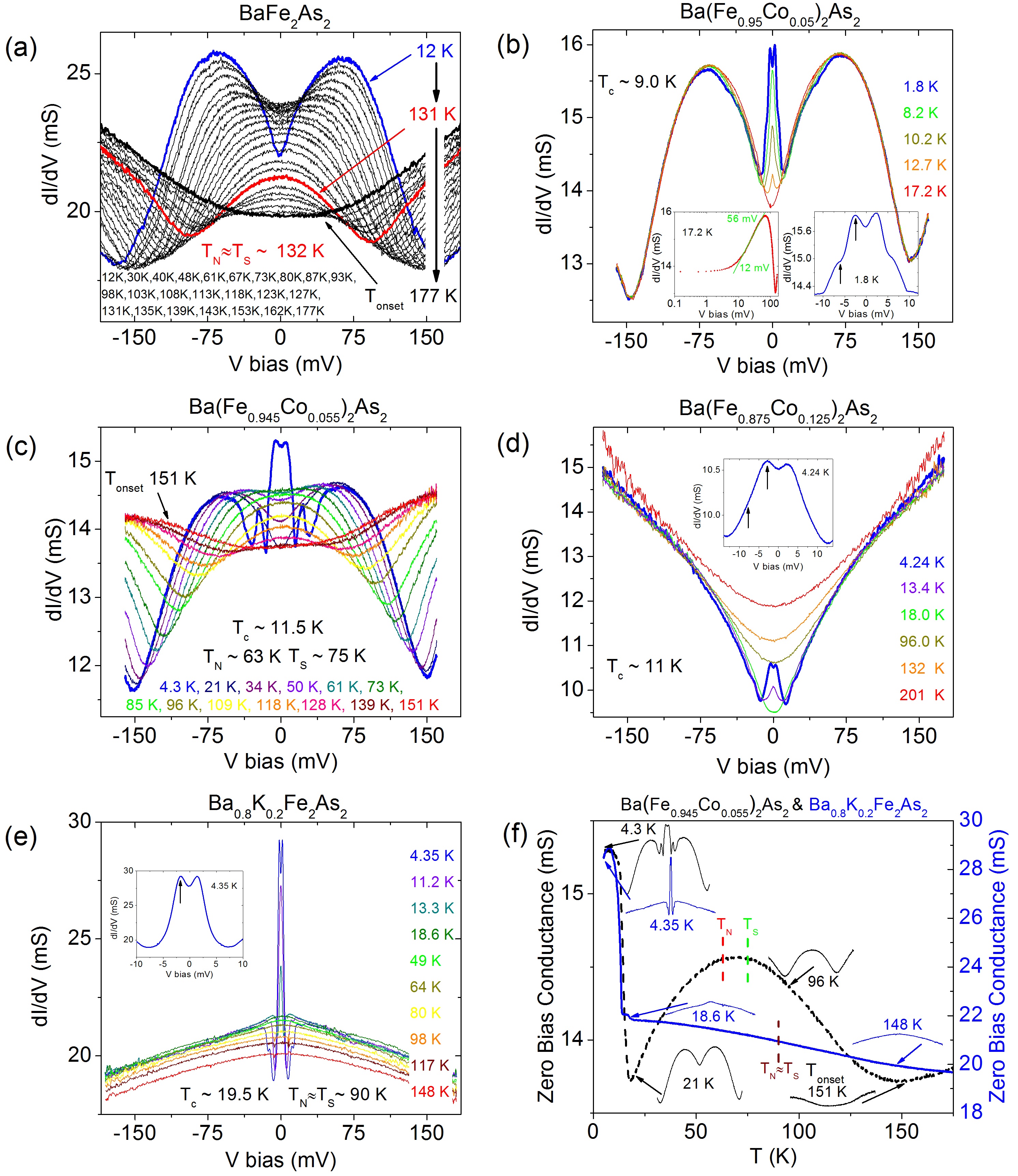}
	\caption{(color online) (a) $G(V)$ for $\rm{BaFe_2As_2}$. Conductance enhancement with peaks at $\sim$ 65mV superimposed on a parabolic background is observed at low temperatures. The peaks move in as the temperature is increased and the enhancement survives well above $T_S$ (red curve). (b, c) $\rm{Ba(Fe_{0.95}Co_{0.05})_2As_2}$ and $\rm{Ba(Fe_{0.945}Co_{0.055})_2As_2}$ display a coexistence of magnetism and superconductivity. At low temperatures, clear Andreev peaks are observed (right inset Fig. 1b, the arrows are pointing out the Andreev peaks). A conductance enhancement with peaks at $\sim$ 65mV coexists with the Andreev spectra and evolves with temperature as it does for $\rm{BaFe_2As_2}$. This enhancement increases logarithmically near zero bias (left inset Fig. 1b). (d) The overdoped compound $\rm{Ba(Fe_{0.875}Co_{0.125})_2As_2}$ shows Andreev spectra below $T_c$. It does not have conductance peaks at higher bias values like the Co underdoped compounds. (e) The hole doped $\rm{Ba_{0.8}K_{0.2}Fe_2As_2}$ has a coexistence of superconductivity and magnetism. It shows Andreev spectra below $T_c$ and no higher bias conductance enhancement. (f) The zero bias conductance (ZBC) vs. temperature curves for $\rm{Ba(Fe_{0.945}Co_{0.055})_2As_2}$ (black dashed curve) and $\rm{Ba_{0.8}K_{0.2}Fe_2As_2}$ (blue solid curve) corresponding to Figs. 1c and e. Both compounds have a magnetostructural transition and a superconducting transition. Co doping shows a conductance enhancement that survives beyond $T_S$ while K doping does not. The insets correlate the spectra obtained at different temperatures to the ZBC curves.}
	\label{fig:100}
\end{figure*}  

\begin{figure*}[htpb]
		\includegraphics[scale=0.7]{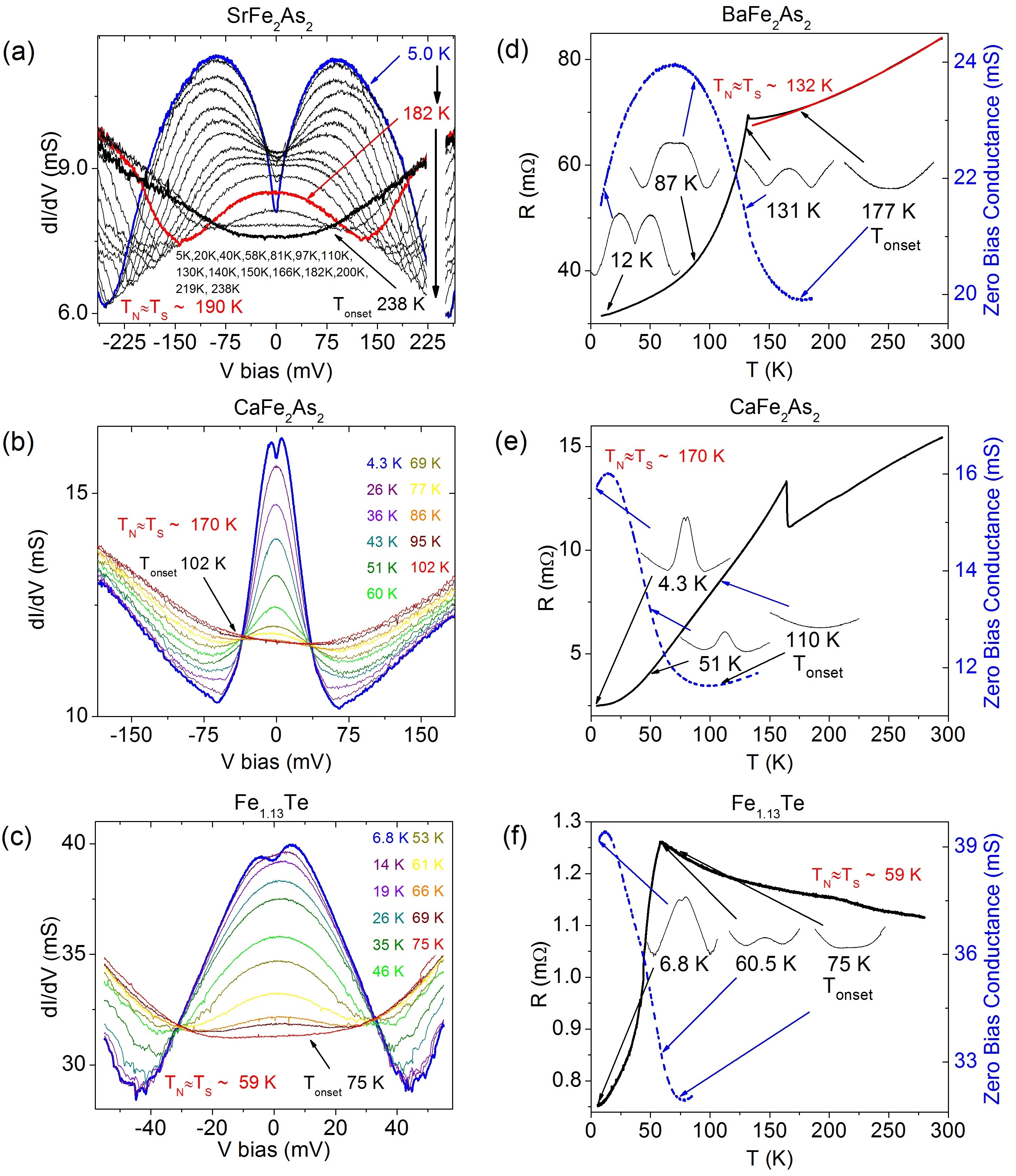}
	\caption{(color online) (a) Conductance spectra for $\rm{SrFe_2As_2}$. The conductance enhancement lasts above $T_S$ and the spectra is similar to that of $\rm{BaFe_2As_2}$ (Fig. 1a). (b) Conductance spectra for $\rm{CaFe_2As_2}$. In this case the conductance enhancement disappears before $T_S$. (c) Conductance spectra for $\rm{Fe_{1.13}Te}$ shows an enhancement that lasts above $T_S$. (d-f) The ZBC (blue dashed curves) and the resistance vs. temperature, (black solid curves) for $\rm{BaFe_2As_2}$, $\rm{CaFe_2As_2}$ and $\rm{Fe_{1.13}Te}$. For $\rm{CaFe_2As_2}$ the enhancement disappears before $T_S$ while for $\rm{BaFe_2As_2}$ and $\rm{Fe_{1.13}Te}$ it lasts into the normal state. The insets correlate the spectra obtained at different temperatures to the ZBC curve. The red curve in (d) is a fit to $\rho=\rho_0+AT^2$.} 
	\label{fig:200}
\end{figure*}  

Fig. 1a shows $G(V)$ for $\rm{BaFe_2As_2}$ ($T_N$ = $T_S$ $\sim$ 132 K, slightly lower than the 134 K reported in literature. \cite{Ni}) At the lowest temperature (blue curve), we see a dip at zero bias and two asymmetric conductance peaks at $\sim$ 65 mV. This double peak feature is superimposed on a parabolic background. (For point contacts on normal metals, at high biases the conductance is slightly downward sloping due to scattering off of non-equilibrium phonons. \cite{Naidyuk} The background observed over here is the opposite.) As the temperature is increased, the dip at zero bias fills, the conductance peaks move inward, and the bias voltage range of the conductance enhancement decreases. No dramatic change in the spectra is observed as $T_S$ is crossed (red curve). The enhancement eventually disappears leaving behind the parabolic background at 177 K, more than 40 K above $T_S$. We define $T_{o}$ as the temperature below which the conductance enhancement is observed. Similar spectra (data not shown) are obtained from two other underdoped non-superconducting samples: x = 0.015 ($T_S$ $\sim$ 119 K, $T_{o}$ $\sim$ 165 K); and x = 0.025 ($T_S$ $\sim$ 107 K, $T_{o}$ $\sim$ 160 K). For Co underdoped samples $T_N$ $<$ $T_S$. 

Fig. 1b shows $G(V)$ for x = 0.05, where long-rage magnetic order coexists with superconductivity ($T_{c}$ $=$ 9.0 K, $T_S$ $\sim$ 78 K). At 1.8 K, the lower bias voltages ($<$ 15 mV) are dominated by Andreev reflection (Fig. 1b inset). However, just like the parent compound, two conductance peaks occur at $\sim$ 65 mV. Above the onset temperature of the superconducting transition, Andreev reflection completely disappears and the high bias conductance evolves just like it does for $\rm{BaFe_2As_2}$. Fig. 1c shows $G(V)$ for another coexisting sample x = 0.055 ($T_{c}$ $=$ 11.5 K, $T_S$ $\sim$ 75 K) depicting the same features. The split Andreev peaks at low temperature attest to the transparency of our junction. The high-bias conductance features are completely reproducible, and are not heating artifacts, as we discuss in Section IV (Non-ideal Point Contact Junctions) of this paper.

Fig. 1d shows $G(V)$ for an overdoped sample with x = 0.125 ($T_{c}$ $=$ 11 K, no $T_S$). At 4.5 K, superconducting Andreev peaks are observed (Fig. 1d inset). Unlike the underdoped samples, no high bias conductance peaks are observed. Above the superconducting transition only a V-shaped background remains. The strength of the V-shaped background varies from junction to junction, and is most likely influenced by the quality of cleaved sample surface. Near optimal doped samples with x = 0.07 ($T_{c}$ $=$ 22 K) and x = 0.08 ($T_{c}$ $=$ 24 K) show similar spectra with Andreev peaks below $T_c$ and a V-shaped background above it (data not shown).

Fig. 1e shows $G(V)$ for $\rm{Ba_{0.8}K_{0.2}Fe_2As_2}$. The sample shows a coexistence of magnetism and superconductivity \cite{Duck1} ($T_N$ = $T_S$ $\sim$ 90 K, $T_{c}$ $\sim$ 20 K). Low temperature spectra show Andreev peaks (Fig. 1e inset). The Andreev signal disappears at $\sim$ 15K leaving behind a background that does not change with a further increase in temperature. Unlike $\rm{BaFe_2As_2}$ and $\rm{Ba(Fe_{1-x}Co_x)_2As_2}$, the background for $\rm{Ba_{0.8}K_{0.2}Fe_2As_2}$ is downward sloping at high bias voltages.  

Fig. 1f shows zero bias conductance (ZBC) curves for $\rm{Ba(Fe_{0.945}Co_{0.055})_2As_2}$ and $\rm{Ba_{0.8}K_{0.2}Fe_2As_2}$. These zero bias conductance curves are for the same junctions whose voltage and temperature evolution has been presented in Figs. 1c and 1e. Superconductivity and magnetism coexist in both samples and both show Andreev spectra below $T_c$. However, while Co doping shows a conductance enhancement that lasts above $T_S$, no such enhancement is observed for K doping. The insets in the figure correlate the spectra obtained at different temperatures to the ZBC curves. The magnetic and structural transition temperatures are marked by vertical dashed lines on the ZBC curves. It is pertinent to note over here that K causing hole doping while Co causes electron doping in $\rm{BaFe_2As_2}$. In addition, Co doping causes the magnetic and structural transitions to split with $T_S > T_N$. For K doping, the two transitions occur at the same temperature ($T_N$ = $T_S$). 

Fig. 2a shows the $G(V)$ for $\rm{SrFe_2As_2}$ and Fig. 2b for $\rm{CaFe_2As_2}$. The trend for $\rm{SrFe_2As_2}$ is very similar to that of $\rm{BaFe_2As_2}$. It has a $T_N$ = $T_S$ of $\sim$ 190K and a $T_{o}$ of $\sim$ 240 K. (Data are taken on an unannealed sample, annealing increases $T_S$ to 200 K).\cite{Gillett} However, $\rm{CaFe_2As_2}$ shows a different behavior. Of the 13 junctions tested, 11 of them showed a conductance enhancement disappearing around 100K-110K, below $T_N$ = $T_S$ $\sim$ 170K. For the remaining 2, the enhancement is observed till 170-180K. 

Fig. 2c shows $G(V)$ for the Fe-chalcogenide $\rm{Fe_{1.13}Te}$. Like $\rm{BaFe_2As_2}$, $\rm{Fe_{1.13}Te}$ shows a conductance enhancement that survives above the magnetic and structural transition temperatures. The conductance enhancement is observed till 75 K ($T_N$ = $T_S$ $\sim$ 59 K). $\rm{Fe_{1.03}Te}$ (data not shown) shows a conductance enhancement that lasts till 85 K ($T_N$ = $T_S$ $\sim$ 69 K). 
 
Figs. 2d-f show ZBC and $R(T)$ curves for $\rm{BaFe_2As_2}$, $\rm{CaFe_2As_2}$ and $\rm{Fe_{1.13}Te}$. They clearly show the enhancement lasting well above $T_S$ for $\rm{BaFe_2As_2}$ and $\rm{Fe_{1.13}Te}$, and disappearing before $T_S$ for $\rm{CaFe_2As_2}$. The insets in the figures correlate the spectra obtained at different temperatures to the ZBC and $R(T)$ curves.  

To summarize thus far, we have studied iron based compounds that exhibit a magnetic and structural transition. $\rm{BaFe_2As_2}$, $\rm{SrFe_2As_2}$, underdoped $\rm{Ba(Fe_{1-x}Co_x)_2As_2}$ and $\rm{Fe_{1+y}Te}$ exhibit a $dI/dV$ enhancement that sets in above $T_S$, $\rm{CaFe_2As_2}$ only shows the enhancement below $T_S$ while $\rm{Ba_{0.8}K_{0.2}Fe_2As_2}$ does not show any conductance enhancement. Overdoped $\rm{Ba(Fe_{1-x}Co_x)_2As_2}$ does not have a $T_S$ and only shows Andreev spectra below $T_c$. The high bias background for all compounds except for $\rm{Ba_{0.8}K_{0.2}Fe_2As_2}$ is an upward facing parabola.  
 
\begin{figure*}[hpbt]
		\includegraphics[scale=0.7]{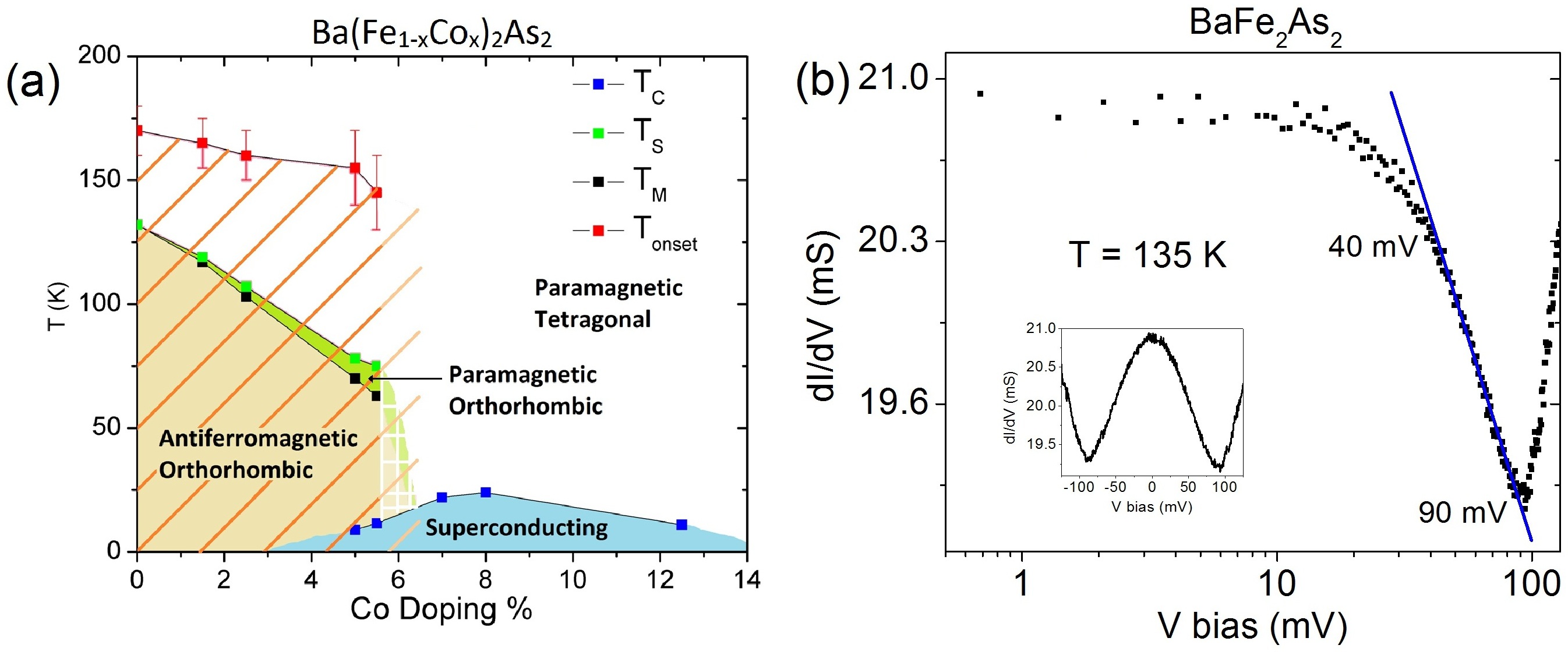}
	\caption{(color online) (a) Phase diagram for $\rm{Ba(Fe_{1-x}Co_x)_2As_2}$. For the underdoped side, a new region is marked (diagonal stripes) indicating the conductance enhancement that sets in above $T_S$. (b) G(V) above $T_S$ for $\rm{BaFe_2As_2}$ follows a log dependence from $\sim$ 40 mV to $\sim$ 90 mV.} 
	\label{fig:311}
\end{figure*}  

\section{Discussion}

We compare the presence of the conductance enhancement with the in-plane resistivity behavior of these compounds. For detwinned underdoped $\rm{AEFe_2As_2}$ it has been shown that below $T_S$ a resistive anisotropy exists.\cite{Tanatar, Blomberg} Above $T_S$ there is notable anisotropy for AE = Ba, neglible anisotroy for AE = Sr and no anisotropy for AE = Ca (Fig. 5 in).\cite{Blomberg} The anisotropy above $T_S$ is sensitive to the uniaxial force required to detwin the samples. Detwinned underdoped $\rm{Ba_{1-x}K_{x}Fe_2As_2}$ does not show any anisotropy at all, either below or above $T_S$. \cite{Ying} In this work we use PCS to probe twinned samples along the c-axis. The presence or absence of the in-plane resistive anisotropy matches with whether a conductance enhancement is detected or not. This indicates that the conductance enhancement we observe is caused by the same phenomenon that causes the in-plane resistive anisotropy in these compounds. Due to inherent difficulties in detwinning $\rm{Fe_{1+y}Te}$, it has not been tested for resistive anisotropy, but since we observe conductance enhancement above $T_S$, we expect it to have a resistive anisotropy that sets in above $T_S$.

The normal state resistivity of metals may be fit to a power law $\rho$ = $\rho_0$+$AT^\alpha$ where $\alpha$ = 2 for standard Fermi liquid theory. Fig. 2d shows such a fit for twinned $\rm{BaFe_2As_2}$. From 300 K down to $\sim$ 180 K, the resistance follows a $T^2$ dependence. The deviation from $\alpha$ = $2$ sets in very close to $T_{o}$. This suggests that the conductance enhancement observed by PCS is tied to the deviation from Fermi liquid behavior in $\rm{BaFe_2As_2}$. 

We construct a revised phase diagram for $\rm{Ba(Fe_{1-x}Co_x)_2As_2}$, (Fig. 3a) marking a new line on the underdoped side showing the temperature below which the conductance enhancement is observed. 

Recent work \cite{Weicheng} has shown that orbital fluctuations above $T_S$ are expected to provide extra contributions to the single particle DOS at zero energy. The DOS follow a log dependence as the energy is increased. Fig. 3b shows that our conductance enhancement for $\rm{BaFe_2As_2}$ above $T_S$ follows a log dependence from $\sim$ 40 mV to $\sim$ 90 mV. Thermal population effects at 135 K causes scatter in the data at low bias voltage. Similar fits are observed above $T_S$ for $\rm{SrFe_2As_2}$ and $\rm{Fe_{1.13}Te}$. Furthermore, the absence of similar effects in our data on $\rm{Ba_{0.8}K_{0.2}Fe_2As_2}$ is consistent with the prediction that crystals that do not show the resistance anisotropy will also not exhibit the excess conductance due to orbital fluctuations.\cite{Weicheng} Our data therefore strongly indicates that the enhancement in conductance observed by our experiments is a consequence of orbital fluctuations. 

It should be kept in mind that the $G(V)$ measured by point contact spectroscopy does not directly correspond to the density of states. Our measured conductance is a convolution of the Fermi velocity and the energy dependent density of states along with any scattering processes that might be present. For normal metals, the Fermi velocity and the density of states are inversely related and cancel each other out. \cite{Harrison} There is a lack of theoretical models for interpreting PCS data on correlated metals, where the DOS are energy dependent and do not cancel out with the Fermi velocity when $G(V)$ is measured. A theory considering both the energy dependence of the electronic DOS and scattering processes would be extremely helpful in obtaining a better understanding of the experimental data.  

With decreasing temperature, the excess conductance curves all develop a dip at zero bias that sharpens as the temperature is lowered further. This could happen if there are two dominant scattering processes with opposite voltage dependence at work, and the crossover between them giving rise to a peak in $G(V)$. PCS on Kondo systems shows a similar effect where the Kondo scattering and phonon scattering give rise to a peak in $G(V)$.\cite{JansenReview} An alternate explanation is that this may be due to the formation of the spin density wave (SDW) gap. Previous work has shown PCS to be sensitive to such gapping.\cite{Meekes,*Escudero} For $\rm{BaFe_2As_2}$, the conductance peak to peak distance lies between 110-140 mV. This agrees well the SDW gap size (100-125 mV) reported by Raman spectroscopy, optical conductivity and ARPES.\cite{Chauvière,*Sugai,Nakajima,*Hu,Richard} However, $G(V)$ increases logarithmically near zero bias (Fig. 1b inset) which lends support to the scattering scenario.  

A maximum is observed at $\sim$ 200K in the interplane c-axis resistivity of $\rm{BaFe_2As_2}$,\cite{Tanatar1} marking the crossover from high temperature nonmetallic to low temperature metallic behavior. The $T_o$ determined from our PCS data occurs at a comparable temperature to this maximum. However unlike our data, the c-axis resistivity maximum is observed at all Co dopings.

Evidence for normal state nematicity from detwinned samples is complicated by the symmetry breaking pressure applied to detwin the crystal. Apart from the resistive anisotropy already discussed, ARPES \cite{Yi} detects orbital ordering, and optical conductivity detects \cite{Dusza} an in-plane anisotropy in the normal state. On twinned samples, inelastic neutron scattering reveals high energy ($>$100meV) spin excitations above $T_S$ in $\rm{BaFe_2As_2}$,\cite{Harriger} although that these are truly indicative of nematicity is unclear.\cite{Kotliar} Torque magnetometry on $\rm{BaFe_2(As_{1-x}P_x)_2}$ detects a $C_4$ symmetry breaking in the normal state, across the phase diagram.\cite{Matsuda1,*Matsuda} Strong anisotropy observed by STM on FeSe, that lacks long range magnetic order, has been explained using orbital ordering.\cite{Song,*Hung}

\section{Non-ideal Point Contact Junctions}

\begin{figure*}[htbp]
		\includegraphics[scale=0.7]{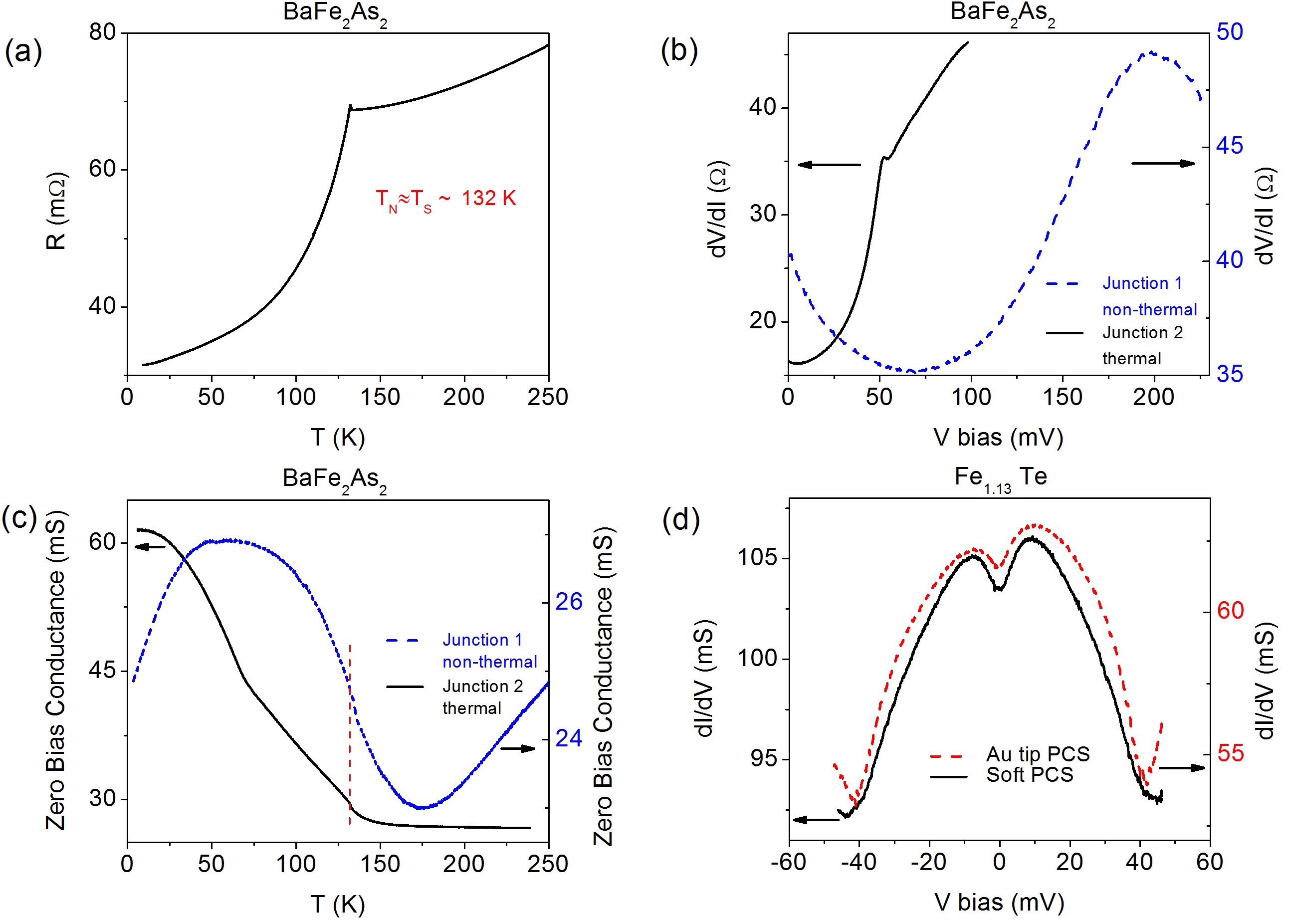}
	\caption{(color online) (a) Resistance vs. Temperature for $\rm{BaFe_2As_2}$. The bulk resistance always rises with an increase in temperature. A gradient change occurs as the magnetostructural transition is crossed. (b) $dV/dI$ for two junctions on $\rm{BaFe_2As_2}$. Junction 2 is in the thermal limit and follows the functional form of the bulk resistivity (black solid curve, taken at 7.6 K). The junction resistance rises with increasing voltage and there is a kink at $\sim$ 52 mV corresponding to being heated across the magnetostructural transition. Junction 1 behaves very differently from bulk resistivity (blue dashed curve, taken at 2.0 K). The junction resistance decreases with an increasing voltage from 0 to $\sim$ 70 mV, and again for voltages larger than $\sim$ 198 mV. A lack of agreement between bulk resistivity and $dV/dI$ indicates that the junction 1 is free of heating effects. (c) The ZBC curve for junction 2 (black solid curve) follows the trend of the bulk resistivity, while that of junction 1 does not (blue dashed curve). The vertical red dashed line marks the magnetostructural transition temperature. (d) Needle-anvil PCS with a Au tip (red dashed curve, taken at 4.0 K) and soft PCS (black solid curve, taken at 6.84 K) on $\rm{Fe_{1.13}Te}$ show very similar spectra.}
	\label{fig:100}
\end{figure*}  

To obtain spectroscopic information from PCS, it is imperative that the junctions are devoid of joule heating effects and any artifacts that may occur due to the junction design. Heating effects will wash out any spectroscopic information while a faulty junction design shall display features not representative of the bulk crystal. In this section we provide evidence that neither is the case for our PCS junctions. We present data on a junction that is in the thermal limit to contrast it with data free of joule heating effects. We also present data obtained by the needle-anvil PCS method to compare it with the data obtained via the soft PCS method.   

\begin{figure*}[hpbt]
		\includegraphics[scale=0.7]{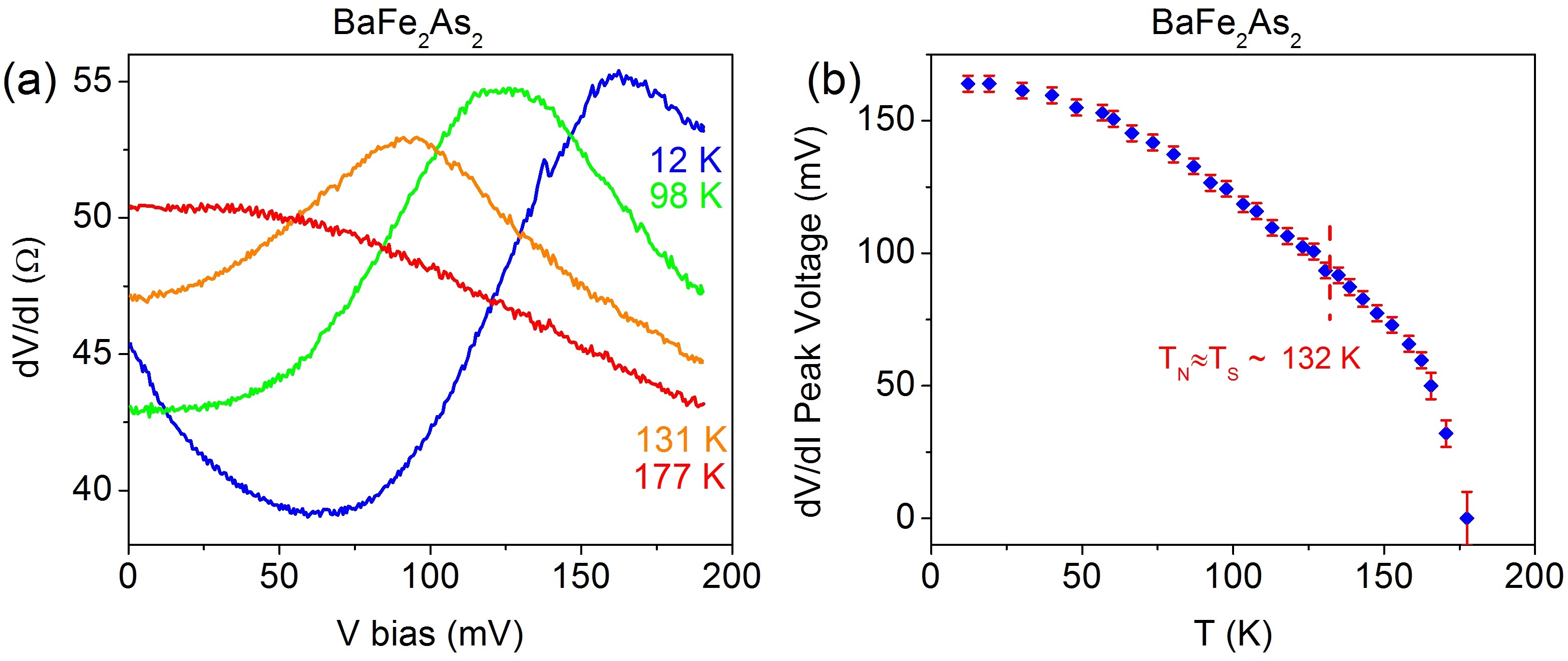}
	\caption{(color online) (a) As the temperature is increased, the peak in $dV/dI$ moves to lower bias voltages. However, it does not disappear at $T_N$ $\sim$ 132 K, and survives until 177 K. (b) The bias voltage at which the peak in $dV/dI$ occurs as a function of temperature. Had the peak been due to heating the junction across $T_N$, it would have disappeared at $T_N$, as shown for thermal limit PCS on magnetic materials by Chien et al. \cite{Chien}}
	\label{fig:200}
\end{figure*}  

\subsection{$dV/dI$ tracks the bulk resistivity in the thermal limit}

For junctions in the thermal regime, there is a strong resemblance between the temperature dependent bulk resistivity $R(T)$ and the voltage dependent point contact resistance $dV/dI$. \cite{Duif} 

Fig. 4 shows $R(T)$ and $dV/dI$ for two soft point contact junctions on $\rm{BaFe_2As_2}$. Both junctions are constructed in the exact same way. First, 50{\AA} $\rm{AlO_x}$ is sputtered on our crystals to act as an insulating barrier and then using Ag paint as a counter electrode, channels are introduced for current transport by fritting \cite{Holm} across the oxide layer. For junction 2 in Fig. 4b (black solid curve), larger channels are opened for current transport as opposed to the junction 1 in Fig. 4b (blue dashed curve), as is evident by their resistances at zero bias (16$\Omega$ vs. 40$\Omega$). Consequently, junction 2 is in the thermal limit while junction 1 shows a $G(V)$ like that shown in Fig. 1a.  

The resistance of $\rm{BaFe_2As_2}$ always increases with rising temperature, and a gradient change occurs at the magnetostructural transition. $dV/dI$ for a junction dominated by heating effects should therefore increase with rising voltage, as a higher voltage would correspond to more heating in the junction. That is exactly what junction 2 in Fig. 4b shows. The kink that occurs at $\sim$ 52 mV corresponds to the junction being heated across the magnetostructural transition. On the other hand, junction 1 shows a completely different scenario. $dV/dI$ actually decreases with an increasing voltage from 0 to $\sim$ 70 mV. There is a bigger chance of heating up the junction at larger bias voltages, but $dV/dI$ starts to decrease again at $\sim$ 198 mV and keeps sloping down till our largest bias value, which is 225 mV for this junction. Similar behavior is observed in $\rm{SrFe_2As_2}$ and $\rm{CaFe_2As_2}$ which are biased up to 270 mV and 200 mV and still show a decreasing $dV/dI$. 

For thermal junctions, the ZBC curve follows a similar functional form to the bulk resistivity. Fig. 4c shows ZBC curves for junctions 1 and 2. For junction 1, the curve is similar to what has already been shown and discussed in Fig. 2d. For junction 2, the ZBC is always decreasing with rising temperature, and there is a gradient change close to the magnetostructural transition, as is observed in the bulk resistivity. The origin of the additional gradient change around 65 K is not clear. The ZBC curves of some joule heated junctions show a greater similarity with the bulk resistivity than others. This is probably determined by how thermal as opposed to diffusive a junction is. 

\subsection{Ideal needle-anvil and soft PCS junctions show similar spectra}
Initial work on this project was performed by using an electrochemically sharpened Au tip to construct junctions on $\rm{Fe_{1+y}Te}$. However, since it is difficult to maintain a stable needle-anvil PCS junction over a wide temperature range, we switched to soft PCS junctions. In order to ensure that our soft point contact junction geometry does not introduce artifacts in our data, we compare the results obtained from the two techniques. Fig. 4d presents spectra on $\rm{Fe_{1.13}Te}$ obtained via needle-anvil PCS and soft PCS. The two methods show very similar curves. This shows that our soft point contact junction setup is not interfering with the intrinsic properties of the material under study. We have also constructed soft PCS junctions on the superconductor $NbSe_2$ and obtained spectra similar to that obtained from needle-anvil PCS.   

\subsection{Thermal limit PCS on magnetic compounds detects the magnetic transition}
There is a tell-tale sign of heating effects for PCS on magnetic compounds, corresponding to their magnetic transition temperature. When the biasing voltage becomes large enough to increase the local temperature of the junction across the magnetic transition temperature, a distinct non-linearity shows up as a turning point in $dV/dI$. \cite{Verkin, Chien} (For the joule heated junction shown in Fig. 4b, this turning point occurs at $\sim$ 52 mV). The turning point occurs at increasingly lower bias voltages as the ambient temperature is increased. Once the ambient temperature is equal to the magnetic transition temperature, the peak in $dV/dI$ occurs at 0 mV. By just looking at our low temperature data for $\rm{BaFe_2As_2}$ in Fig. 1a, it may be speculated that the maximum in $dV/dI$ at $\sim$ 162 mV is the sample crossing the magnetic transition. However, had that been the case, this maximum would have disappeared for $dV/dI$ curves taken at $T>T_N$. As Fig. 5a shows, the peak in $dV/dI$ is present at $T_N$ and has moved inwards to 94 mV. It eventually disappears at 177 K. Therefore, it cannot possibly correspond to the junction getting warm enough to cross the magnetic transition. We also plot the peak position as a function of temperature in Fig. 5b.

\begin{figure*}[hpbt]
		\includegraphics[scale=0.7]{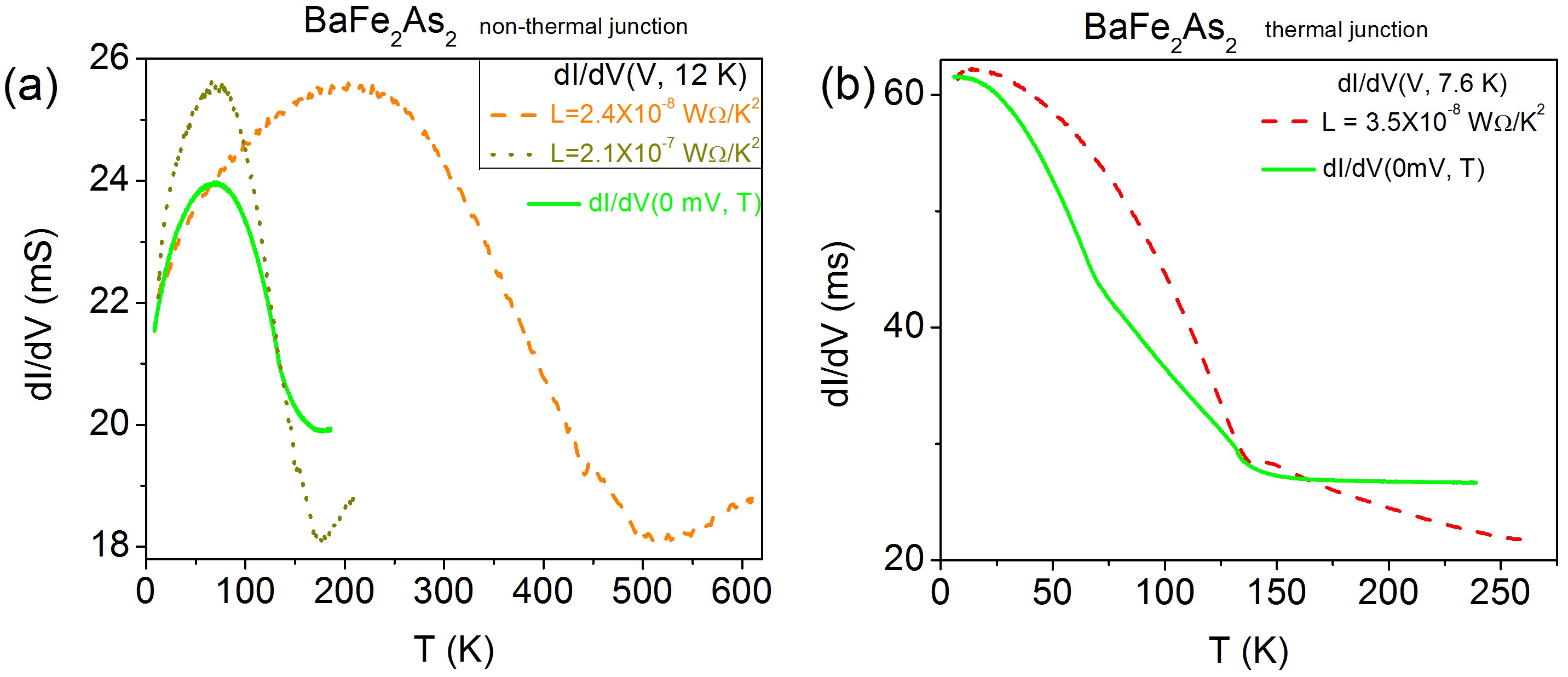}
	\caption{(color online) (a) Comparing the ZBC and $dI/dV$ curves for $\rm{BaFe_2As_2}$. The orange dashed and dark yellow dotted curves are the low temperature $dI/dV$ curve plotted on a temperature scale using $T_{max}^2=T_{bath}^2+V^2/4L$. The orange dashed curve uses L = 2.44 x $10^{-8}W \Omega K^{-2}$ while the dark yellow dotted curve uses L = 2.1 x $10^{-7}W \Omega K^{-2}$. A lack of agreement between the ZBC (solid green curve) and these extrapolated $dI/dV$ curves shows that the junction is not in the thermal limit. (b) Comparing the ZBC and $dI/dV$ curves for a joule heated junction on $\rm{BaFe_2As_2}$. The kink in the $dI/dV$ spectrum occurs at the same conductance value at which the ZBC crosses the magnetostructural transition, and the two curves may be made to overlap using L = 3.5 x $10^{-8}W \Omega K^{-2}$. The lack of a fit at lower bias voltages/lower temperatures implies that either L is varying with temperature or the junction is not completely in the thermal regime.} 
	\label{fig:311}
\end{figure*}  

\subsection{Converting $dI/dV$(V, low T) into zero bias conductance $dI/dV$(0 mV, T) by using the Lorentz number}
Another check for thermal PCS junctions is to compare the $dI/dV$(V, low T) curve with the zero bias conductance curve, $dI/dV$(0 mV, T). The local temperature in a thermal junction is related to the bias voltage by $T^2_{max} = T^2_{bath} + V^2/4L$ where L is the Lorentz number of the compound. \cite{Duif} In such a scenario, there is a substantial overlap between ZBC and $dI/dV$, as shown for $UPt_3$ in.\cite{Duif} For ballistic junctions, ZBC and $dI/dV$ may superficially have the same shape but they will no longer have any overlap (for e.g. Fig. 1 in).\cite{Jansen3} In Fig. 6a we compare these two quantities for the $\rm{BaFe_2As_2}$ junction whose temperature evolution was shown in Fig. 1a. The orange dashed curve uses the value of L for $\rm{BaFe_2As_2}$ reported in \cite{Kurita}; L = 2.44 x $10^{-8}W \Omega K^{-2}$. Since the L reported in \cite{Kurita} is for low temperature and L may vary with temperature, we also compare $dI/dV$ and ZBC for L = 2.1 x $10^{-7}W \Omega K^{-2}$, that forces 170 mV to correspond to 175 K (dark yellow dotted curve). There is still no quantitative agreement between  $dI/dV$ and ZBC since the maximum and minimum values for the two curves are quite different. The lack of any overlap for either value of L shows that our junction is not in the thermal limit. In Fig. 6b we compare the $dI/dV$ curve with the ZBC curve for junction 2. The kink at the magnetostructural transition occurs at the same conductance value in both curves, providing evidence for the thermal nature of the junction. Voltage has been converted temperature using L = 3.5 x $10^{-8}W \Omega K^{-2}$. The lack of a fit at lower bias voltages/lower temperatures implies that either L is varying with temperature or the junction is not completely in the thermal regime. 

The striking similarity between the $dI/dV$ and the ZBC curves for junctions not impacted by heating effects implies that the scattering processes responsible for our spectra behave in a similar manner under temperature and voltage. An existing example of this is PCS on Kondo systems, where the scattering has a log dependence on both temperature and voltage.\cite{JansenReview} Fig. 1 in Fisun et al. \cite{Fisun} shows that for ballistic PCS on Kondo systems, $dV/dI$ and R(T) have the same functional form.   

\subsection{Different bulk resistivity but similar dI/dV spectra for Fe-pnictides vs. Fe-chalcogenides}
The Fe-pnictides and the Fe-chalcogenides show very different resistivity curves (Figs. 2d-f) and may be classified as bad metals.\cite{Johnston} Upon cooling the $\rm{Fe_{1.13}Te}$ from room temperature to $T_S$, the resistance is observed to increase, while for the $\rm{BaFe_2As_2}$, the resistance is seen to decrease. Below $T_S$, the resistance decreases much more quickly with decreasing temperature for $\rm{Fe_{1.13}Te}$ than for $\rm{BaFe_2As_2}$. Despite differences in their $R(T)$, the two families show similarly shaped $G(V)$ spectra. This is further evidence that our junctions are not in the thermal regime and that the same scattering mechanisms are at work in both the Fe-chalcogenides and the Fe-pnictides. 

To summarize, via soft PCS, we can construct junctions both highly impacted by joule heating effects and junctions free of heating effects. All the data presented and discussed in this paper has been taken on junctions free of joule heating. There is ample evidence that these junctions are not in the thermal limit: (a) No agreement of bulk resistivity with $dV/dI$ (b) No indication that the low temperature curves cross the magnetic transition when biased to high voltages (c) No quantitative agreement between the $dI/dV$ and the ZBC curves (d) $G(V)$ spectra of similar functional form obtained from the Fe-chalcogenides and Fe-pnictides, who have different temperature dependence of bulk resistivity. In addition, since Au tip and soft PCS junctions show similar spectra, our features are not an artifact of our junction fabrication technique.

\section{Summary}

In conclusion, we have performed PCS measurements on twinned $\rm{AEFe_2As_2}$ (AE = Ca, Sr, Ba), $\rm{Ba(Fe_{1-x}Co_x)_2As_2}$, $\rm{Ba_{0.8}K_{0.2}Fe_2As_2}$ and $\rm{Fe_{1+y}Te}$. For superconducting samples at low biases ($<$ 15 mV), we observe Andreev spectra with split conductance peaks. We also observe a conductance enhancement which is indicated to arise from orbital fluctuations that increase the single particle DOS at zero energy.\cite{Weicheng} The enhancement sets in above $T_S$ for those compounds that are known to exhibit an in-plane resistive anisotropy above $T_S$. Based on our PCS data, therefore, an in-plane resistive anisotropy of untwined  $\rm{Fe_{1+y}Te}$ that onsets above $T_S$ is anticipated. At low temperatures, this conductance enhancement develops a dip at zero bias for reasons that are not entirely clear. Our experiments provide support for orbital nematicity in $\rm{BaFe_2As_2}$, $\rm{SrFe_2As_2}$ and $\rm{Fe_{1+y}Te}$. We also construct point contact junctions that are in the thermal limit. These junctions are impacted by joule heating and follow bulk resistivity effects.

It is worth reiterating that the experimentally measured $G(V)$ is not a direct representation of the electronic DOS but rather a convolution of the DOS with the Fermi velocity and any other scattering mechanisms that may be present. A theory that clarifies the relationship between the $G(V)$ measured by PCS on the iron based compounds and the electronic DOS shall be of great help in further explanation and analysis of our data.

\begin{acknowledgments}

We thank P. Phillips and W.-C. Lee for valuable and motivating discussions. We also thank N. Ni for sharing resistance data for $\rm{Ba(Fe_{1-x}Co_x)_2As_2}$ and Z. X. Shen for sharing ARPES data for $Fe_{1+y}Te$. This work is supported as part of the Center for Emergent Superconductivity, an Energy Frontier Research Center funded by the US Department of Energy, Office of Science, Office of Basic Energy Sciences under Award No. DE-AC0298CH1088. Ames Lab is operated by ISU under DOE Contract No. DE-AC02-07CH11358. University of Cambridge is supported by EPSRC, Trinity College, the Royal Society and the Commonwealth Trust.

\end{acknowledgments}

\bibliography{myrefs}
\end{document}